\pdfoutput=1
\documentclass[10pt]{article}
\usepackage{amsmath,amssymb}
\usepackage{graphicx}
\usepackage{color}
\usepackage{float}
\usepackage{caption}
\usepackage{makecell}
\usepackage{textcomp} 
\usepackage{listings}
\usepackage[backend=biber,bibencoding=utf8, style=numeric,sorting=none]{biblatex}
\usepackage[top=1in, bottom=1in, left=1in, right =1in]{geometry}
\bibliography{references}
\usepackage{breqn}
\usepackage[affil-it]{authblk}
\usepackage{abstract}
\usepackage[caption=false]{subfig}
\usepackage{tabularx}
\usepackage{hyperref}
\hypersetup{
    unicode=false,          
    pdftoolbar=true,        
    pdfmenubar=true,        
    pdffitwindow=false,     
    pdfstartview={FitH},    
    pdfauthor={Anshuman Baruah},     
    colorlinks=true,       
    linkcolor=blue,          
    citecolor=blue,        
}

\makeatletter
\newbox\abstract@box
\renewenvironment{abstract}
  {\global\setbox\abstract@box=\vbox\bgroup
     \hsize=\textwidth\linewidth=\textwidth
    \small
    \begin{center}%
    {\bfseries \abstractname\vspace{-.5em}\vspace{\z@}}%
    \end{center}%
    \quotation}
  {\endquotation\egroup}
\expandafter\def\expandafter\@maketitle\expandafter{\@maketitle
  \ifvoid\abstract@box\else\unvbox\abstract@box\if@twocolumn\vskip1.5em\fi\fi}
\makeatother
\begin{document}
\definecolor{dkgreen}{rgb}{0,0.6,0}
\definecolor{gray}{rgb}{0.5,0.5,0.5}
\definecolor{mauve}{rgb}{0.58,0,0.82}

\lstset{frame=tb,
  	language=Matlab,
  	aboveskip=3mm,
  	belowskip=3mm,
  	showstringspaces=false,
  	columns=flexible,
  	basicstyle={\small\ttfamily},
  	numbers=none,
  	numberstyle=\tiny\color{gray},
 	keywordstyle=\color{blue},
	commentstyle=\color{dkgreen},
  	stringstyle=\color{mauve},
  	breaklines=true,
  	breakatwhitespace=true
  	tabsize=3
}


\title{NEW WORMHOLE SOLUTIONS IN A VIABLE $f(R)$ GRAVITY MODEL}
\author{Anshuman Baruah \thanks{E-mail: \textsf{anshuman.baruah@aus.ac.in}}}
\author{Parangam Goswami}
\author{Atri Deshamukhya}
\affil{Department of Physics, Assam University, Silchar, Assam, India-788011}
\date{}

\begin{abstract}
Traversable wormhole solutions in General Relativity require \emph{exotic} matter sources that violate the null energy condition (NEC), and such behavior maybe avoided in modified gravity. In this study, we analyze the energy conditions for static, spherically symmetric traversable Morris-Thorne wormholes in a recently proposed viable $f(R)$ gravity model. We numerically analyze solutions considering both constant and variable redshift functions, and present wormhole space-times respecting the NEC, supported by a phantom energy-like equation of state for the source. Moreover, we analyze the stability of the space-times using the generalized Tolman-Oppenheimer-Volkov equation. We demonstrate the effects of certain parameters in the $f(R)$ model in determining energy condition violations, and establish that stable wormholes can be formulated only at the expense of violating the NEC.
\end{abstract}
\maketitle
\section{\label{sec:intro}Introduction}
Wormholes are compact regions of space-time with topologically simple boundaries and topologically non-trivial interiors \cite{Visser:1995cc}. They are solutions to the Einstein's field equation (EFE) of General Relativity (GR), describing two asymptotically flat regions of space-time connected by a `throat'. Both asymptotically flat surfaces may be separated by arbitrarily large distances. Stable wormhole solutions can be formulated in GR \cite{Morris:1988cz} at the expense of violating the null energy condition (NEC), which ensures that the observed energy density of matter fields in the EFEs is positive for any observer. Traversable wormholes in GR can be described by the following line element \cite{Morris:1988cz} :
\begin{equation}
     ds^2 = - e^{2 \Phi(r)} dt^2 + \frac{dr^2}{ 1 - \frac{b(r)}{r}} + r^2 {d{\theta}}^2 + r^2 sin^2 {\theta} d {\phi}^2,
     \label{mtle}
\end{equation}
where $\Phi(r)$ determines the gravitational redshift, and is called the redshift function; and $b(r)$ determines the spatial shape of the wormhole, and is called the wormhole shape function. Traversable wormholes allow signals that fall in an asymptotically flat region to emerge through the throat (located at some $r=r_0$) in a different asymptotically flat region. In order for the wormhole described by Eq. \eqref{mtle} to be traversable, the following constraints are imposed on the metric functions $\Phi(r)$ and $b(r)$. Firstly, event horizons are not allowed in wormhole space-times. Horizons in spherically symmetric space-times are identified by physically non-singular surfaces at $g_{00}=-e^{2\Phi}\rightarrow 0$, which leads to the constraint that $\Phi(r)$ must be finite everywhere throughout the space-time. Secondly, the throat of the wormhole is located at some minimum radius $r=r_0$, where $b(r_0) = r_0$. Since the corresponding metric component is singular at the throat, the proper distance $l(r)$, defined in Eq. 2, must be well-behaved.
\begin{equation}
    l(r)= \pm \int_{r_0}^r \frac{dr}{\sqrt{1-b(r)/r}}
\end{equation}
Thus, the above integral must be regular throughout the space-time, which requires that $1-b(r)/r \geq 0$. Moreover, $b(r)$ should satisfy the \emph{flaring-out} condition \cite{Morris:1988cz} given by $(b(r) - b'(r)r)/b^2 > 0$.
These constraints on the metric functions impose constraints via the EFEs on the mass-energy density and radial and lateral pressures of the matter that threads the geometry, which lead to violations of various energy conditions \cite{Morris:1988cz}. The matter source threading such wormholes is conventionally described by an anisotropic perfect fluid with a diagonal stress-energy tensor, $T_{\mu\nu} = \text{diag}(\rho, p_r, p_t, p_t)$,
where $\rho$ is the energy density, $p_r$ is the radial pressure, and $p_t$ is the lateral pressure. The NEC requires that $\rho+p_r \geq 0$, whereas wormholes in GR require that $\rho+p_r < 0$ for matter sources threading wormhole space-times \cite{Morris:1988cz}. Moreover, this leads to the violation of the weak energy condition (WEC), $\rho \geq 0$ (\emph{exotic} matter). 
\\
Several modifications to Einstein's GR have been proposed to address its shortcomings. For example, observations of type-Ia supernovae and the cosmic microwave background have confirmed the late-time accelerated expansion of the Universe. In the $\Lambda$-CDM model of cosmology based on GR, additional fields are required to describe the accelerated expansion of the Universe (the \emph{dark energy} problem). However, in modified gravity theories, late-time cosmic accelerated expansion can be realized without any additional fields. Since traversable wormholes in GR cannot be sourced by known matter sources, a natural question that arises is can wormholes without exotic matter be realized in modified theories of gravity.
\\
The gravitational field equations in modified/extended gravity can be recast as:
\begin{equation}
    g_1 (\Psi^i) (G_{\mu \nu} + H_{\mu \nu}) = 8 \pi g_2 ( \Psi^j )T_{\mu \nu}
\end{equation}
where $H_{\mu \nu}$ comprises geometrical corrections to GR in modified gravity, $g_i (\Psi^i)$ are multiplicative factors, and $\Psi^i$ are curvature invariants or other fields contributing to the dynamics of the theory. Accurately identifying $g(\Psi^j) $ and $H_{\mu \nu}$ makes it possible to formulate wormhole solutions in a manner such that the matter stress energy obeys the corresponding NEC ($T_{\mu \nu}k^\nu k^\nu \geq 0$), and violations may be attributed to additional curvature terms. Additionally, Capozziello et al. \cite{capo2, capozziello2015} considered the diffeomorphism invariance of the matter action and contracted Bianchi identities to derive generalized energy conditions in modified gravity. For example, the generalized NEC for $f(R)$ gravity can be written as $T_{\mu \nu} k^\mu k^\nu + k^\mu k^\nu \frac{\nabla_\mu \nabla_\nu F}{F} \geq 0$, with $F \equiv d f(R)/dR$. In such a setting, an appropriate form of $f(R)$ may be considered such that the above inequality holds. Thus, owing to the inherently different structure of the field equations in modified gravity, phenomena such as cosmic acceleration and wormholes may be realized with matter sources satisfying the energy conditions, and violations attributed to additional degrees of freedom in the theory.  
To this end, traversable wormholes have been studied extensively in $f(R)$ modified gravity theory \cite{Tiberiu2013, PhysRevD.80.104012, debenedictis2012wormhole, 2004}, which is perhaps the most extensively discussed modification of GR \cite{2010felice, review1}. $f(R)$ theories are also of significant interest in cosmology with regard to inflation and late-time acceleration of the observable Universe \cite{1979JETPL,STAROBINSKY198099,staro,2007sawicki,2008tsuji, moraes2019phantom}, of late, in the context of gravitational waves \cite{2019gw}. Recently, traversable wormhole solutions have been studied extensively in different $f(R)$ gravity theories with different choices of metric parameters \cite{2019,2019A,2019B,2019C,2019D,doi:10.1142/S0218271820500686,2020A,2020B,2020C}. Wormholes supported by ordinary matter in $f(R)$ gravity have also been reported previously \cite{doi:10.1142/S0217732316501923, doi:10.1142/S0217732311035547}. In this study, we focus on a recently proposed $f(R)$ gravity model \cite{PhysRevD.77.046009, Gogoi:2020ypn}, and analyze the energy conditions for Morris-Thorne wormhole solutions. We explore possible solutions with constant and variable redshift, and demonstrate that traversable wormholes satisfying the energy conditions can be obtained in this theory with suitable choices of parameters. However, our results indicate that such space-times are unstable. Using the generalized Tolman-Oppenheimer-Volkov (TOV) equation, we show that stable wormholes in this $f(R)$ model will violate the NEC, and some model parameters play a crucial role in determining these violations.
\\
The remainder of this manuscript is organized as follows: in Sec. 2, we describe the basic features of Morris-Thorne wormholes in $f(R)$ gravity, introduce the new model studied, and setup the field equations. In Sec. 3, we present our main results, and conclude the manuscript in Sec. \ref{sec:conc}. Throughout the manuscript, we adhere to natural units with $G=c=1$.

\section{The Morris-Thorne Wormhole in $f(R)$ gravity}
\label{sec:whgeometry}
In $f(R)$ gravity, the standard Einstein-Hilbert action takes the following form:
\begin{equation}\label{fraction}
    S = \int d^4 x \sqrt{-g} \left[ f(R) + \mathcal{L}_m \right]
\end{equation}
where $f(R)$ is a function of the Ricci scalar, and $\mathcal{L}_m$ collectively denotes the matter Lagrangian for possible matter fields. In the metric formalism, the following field equation gives the modified EFE for $f(R)$ gravity:
\begin{equation}
FR_{\mu\nu}-\frac{1}{2}f(R)g_{\mu\nu}-\nabla_\mu \nabla_\nu
F+g_{\mu\nu}\Box F=T^m_{\mu\nu},
\label{eq:3}
\end{equation}
where $F\equiv df(R)/dR$, and $T^m_{\mu\nu} = \frac{-2}{\sqrt{-g}} \frac{\delta \mathcal{L}_m}{\delta g^{\mu \nu}}$ denotes the matter stress-energy tensor. We assume that $g_{\mu\nu}$ describes a spherically symmetric space-time described by the line-element in Eq. \eqref{mtle}. The field equation Eq. \eqref{eq:3} can be contracted to yield the following:
\begin{equation}
FR-2f(R)+3\Box F=T
\end{equation}
 \label{kionulai}
Here, $T$ is the trace of the matter stress energy tensor and $\Box F$ is given by the following expression, with $F'=d f(R)/d R$ and $b'=d\,b(r)/dr$:
\begin{align}
    \Box F = \frac{1}{\sqrt{-g}} \partial_\mu (\sqrt{-g} g^{\mu \nu} \partial_\nu F) = \left(1-\frac{b}{r}\right)\left[F''
-\frac{b'r-b}{2r^2(1-b/r)}\,F'+\frac{2F'}{r}\right]
\end{align}
It can be verified that GR (without a cosmological constant) is recovered by setting $f(R)=R$ in Eq. \eqref{fraction}, which implies $F = 1$, and $\Box F=0$. However, $\Box F \neq 0$ in modified gravity theories, and it is generally interpreted as a propagating scalar degree of freedom. Now, from Eq. \eqref{eq:3} we have the following modified EFE:
\begin{equation}
G_{\mu\nu}\equiv R_{\mu\nu}-\frac{1}{2}g_{\mu\nu} R= T^{{\rm
eff}}_{\mu\nu} = T^{{\rm c}}_{\mu\nu} + \frac{T^m_{\mu\nu}}{F}
    \label{field:eq2}
\end{equation}
Here, $T^{{\rm eff}}_{\mu\nu}$ is an effective stress-energy tensor, which may be interpreted as a gravitational fluid responsible for NEC violations in wormhole solutions. It contains the matter stress energy tensor $T^m_{\mu\nu}$ and curvature stress-energy tensor $T^{{\rm c}}_{\mu\nu}$ given by:
\begin{equation}
T^{c}_{\mu\nu}=\frac{1}{F}\left[\nabla_\mu \nabla_\nu F
-\frac{1}{4}g_{\mu\nu}\left(RF+\Box F+T\right) \right]
    \label{gravfluid}
\end{equation}
The basic geometric requirements for wormhole space-times have been discussed in the previous section. We aim to investigate the properties of such a space-time within the framework of $f(R)$ gravity. We consider that $T^m_{\mu \nu}$ describes an anisotropic distribution of matter threading the wormhole geometry, described by the following diagonal stress-energy tensor: 
\begin{equation}
T^m_{\mu\nu}=(\rho+p_t)U_\mu \, U_\nu+p_t\,
g_{\mu\nu}+(p_r-p_t)\chi_\mu \chi_\nu \,,
\end{equation}
where $U^\mu$ is a four-velocity, and $\chi^\mu$ is a unit space-like vector. Now, using the line element in Eq. \eqref{mtle}, the effective field equation Eq. \eqref{field:eq2} can be used to obtain the following general expressions for the matter fields following the treatment in \cite{PhysRevD.80.104012}:
\begin{eqnarray}
\label{generic1} \rho&=&\frac{Fb'}{r^2}\,
       \\
\label{generic2}
p_r&=&-\frac{bF}{r^3}+\frac{F'}{2r^2}(b'r-b)-F''\left(1-\frac{b}{r}\right)
     \,   \\
\label{generic3}
p_t&=&-\frac{F'}{r}\left(1-\frac{b}{r}\right)+\frac{F}{2r^3}(b-b'r)\,
\end{eqnarray}
Note that in the above equations, it has been assumed that $\Phi(r)$ is constant throughout the space-time (i.e., $\Phi'(r) = 0$). Thus, we have $R= \frac{2b'}{r^2}$ for the metric \eqref{mtle}. Such space-times without tidal forces have been studied intensively in literature, and the main motivation arises from the fact that setting $\Phi'(r)=0$ simplifies the calculations significantly. However, the redshift function plays a crucial role in determining the behavior of the energy condition inequalities for wormholes, and assuming a constant $\Phi(r)$ may be an over simplification. Therefore, we also consider wormhole solutions with a non-constant redshift function in our analyses. With $\Phi'(r) \neq 0$, we have 

\begin{equation}
    R=\left(1-\frac{b(r)}{r}\right) \left(2 \Phi ''(r)+4 \Phi '(r)^2+\frac{4 \Phi '(r)}{r}\right)-\frac{r b'(r)-b(r)}{r^2}\Phi '(r)-\frac{2 b'(r)}{r^2}
\end{equation}

Then, we have the following general expressions:
\begin{equation}\label{vrsfone}
    \rho=\frac{Fb'(r)}{r^2}-\Bigg(1-\frac{b(r)}{r}\Bigg)F'\Phi'(r)-H
\end{equation}
\begin{align}\label{vrsftwo}
    p_r=-\frac{b(r)F}{r^3}+2\Bigg(1-\frac{b(r)}{r}\Bigg)\frac{\Phi^{'}(r)F}{r} -\Bigg(1-\frac{b(r)}{r}\Bigg)\Bigg[F''+\frac{F'(rb'(r)-b(r))}{2r^2\Big(1-\frac{b(r)}{r}\Big)}\Bigg]+H
\end{align}
\begin{align}\label{vrsfthree}
p_t=\frac{F(b(r)-rb'(r))}{2r^3}-\frac{F'}{r}\Bigg(1-\frac{b(r)}{r}\Bigg)
&+F\Bigg(1-\frac{b(r)}{r}\Bigg)\\ \nonumber\Bigg(\Phi^{''}(r)-\frac{(rb'(r)-b(r))\Phi^{'}(r)}{2r(r-b)}  &+\Phi^{'2}(r)+\frac{\Phi^{'}(r)}{r}\Bigg)+H
\end{align}

where $H(r)=\frac{1}{4}\left(FR+\Box F +T\right)$ for notational simplicity. The above system of equations comprises the geometric parameters $b(r)$ and $\Phi(r)$, the thermodynamic variables $\rho$, $p_r$, and $p_t$, and the $F(r)$\footnote{The functional $F(R(r)$) has been denoted as $F(r)$ for notational simplicity} functional. The system of equations can be closed mathematically by specifying three of these parameters. It is worth noting that although it is possible to solve for the functional $F(r)$ by fixing $b(r)$, the $f(R)$ obtained via integration may not be physically interesting. Thus, we leverage a recently proposed cosmologically viable $f(R)$ gravity model \cite{PhysRevD.77.046009, Gogoi:2020ypn} with:
\begin{equation}\label{gog}
    f(R) = R-\frac{\alpha}{\pi } R_c \cot ^{-1}\left(\tfrac{R_c^2}{R^2}\right) -\beta  R_c \left[1-\exp\left({-\tfrac{R}{R_c}}\right)\right]
\end{equation}
where $\alpha$ and $\beta$ are dimensionless constants, and $R_c$ is a characteristic curvature constant. This theory fulfills the basic requirements for a viable $f(R)$ theory \cite{PhysRevD.90.103512,2008tsuji} for suitable choices of $\alpha$, $\beta$, and $R_c$, including solar system tests. Moreover, the model has been investigated thoroughly in terms of observational constraints such as gravitational wave polarization modes and recent data from pulsar timing arrays \cite{Gogoi:2020ypn}, which may have important implications in certain astrophysical and cosmological problems. \\
Since we are mainly concerned with the energy condition inequalities in the framework of viable $f(R)$ gravity, we can fix $b(r)$ and $\Phi(r)$ in the system of equations for the analysis. The geometric requirements on the space-time and asymptotic behavior can be satisfied by constraining these functions appropriately.
It is worth highlighting here that a novel approach of formulating the shape function in a model-independent manner has recently been proposed\cite{capozziello2021traversable}, which may also serve as a suitable approach to investigate or constrain NEC violations. To further probe the nature of the wormhole space-times, we examine the equation of state (EoS) parameter, $\omega = p_r/ \rho$, and anisotropy parameter, $\Delta = p_t - p_r$, for possible wormhole solutions. Further, we check the stability of the solutions using the generalized TOV equation \cite{PhysRev.55.374,tov2,tov3}.
In our analyses, we consider the following two cases:
\begin{enumerate}
    \item Case I: Constant $\Phi(r)$ with $b (r)=\frac{r_0 \log (r+1)}{\log (r_0+1)}$
    \item Case II: $\Phi(r) =\sqrt{\frac{r_0}{r}}$ and $b (r)= r_0 \log \left(\frac{r}{r_0}\right)+r_0$
\end{enumerate}
The explicit forms of $\rho$, $p_r$, and $p_t$ for Cases I and II are rather lengthy, and have been included in \ref{AppendixB}.

\section{Results and Discussion}\label{sec:res}
Fig. 1 shows the profile of $\rho$ against $r$ for different combinations of model parameters in arbitrary units, with $R_c=1$ and $r_0=0.9$ for Cases I and II, respectively. It can be seen that the energy density is positive (WEC holds) at the throat in Case I, while in Case II, the WEC holds at the throat with $\alpha = 0.15, \beta=0.50$ (we denote this as \textit{Model 1} hereafter).
\begin{figure}[H]
   \includegraphics[width=\columnwidth]{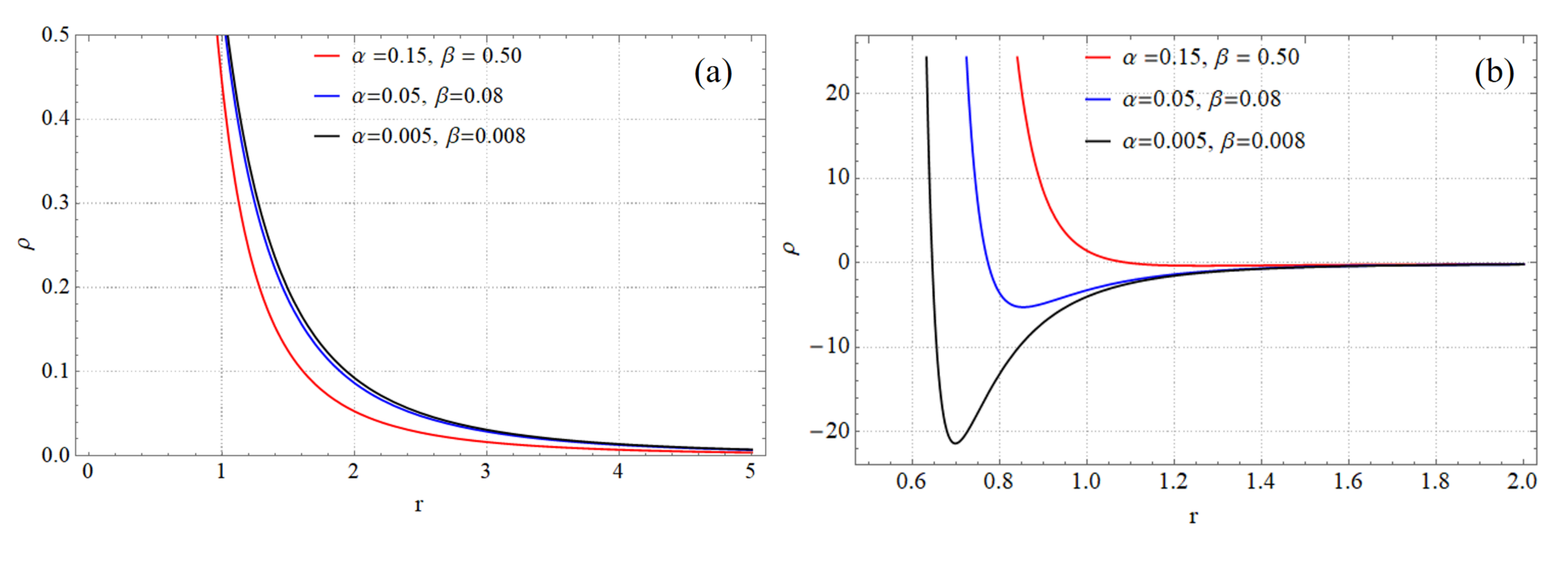}
   \caption{Profile of the energy density for different combinations of model parameters in arbitrary units with $R_c=1$ for (a) Case I (constant redshift function) and (b) Case II (variable redshift function).}\label{two}
\end{figure}
Secondly, we checked the NEC terms for both cases. In Case I, $\rho+p_r < 0$ at the throat, and $\rho+p_t > 0$, as shown in Figs. 2(a) and 3(a). Since both these terms should be $>0$ at the throat, we conclude that the NEC is violated in Case I. In Case II,  $\rho+p_r = 0$, and $\rho+p_t > 0$ only for \textit{Model 1}, as shown in Figs. 2(b) and 3(b). Thus, the NEC is satisfied for \textit{Model 1}.

\begin{figure}[H]
   \includegraphics[width=\columnwidth]{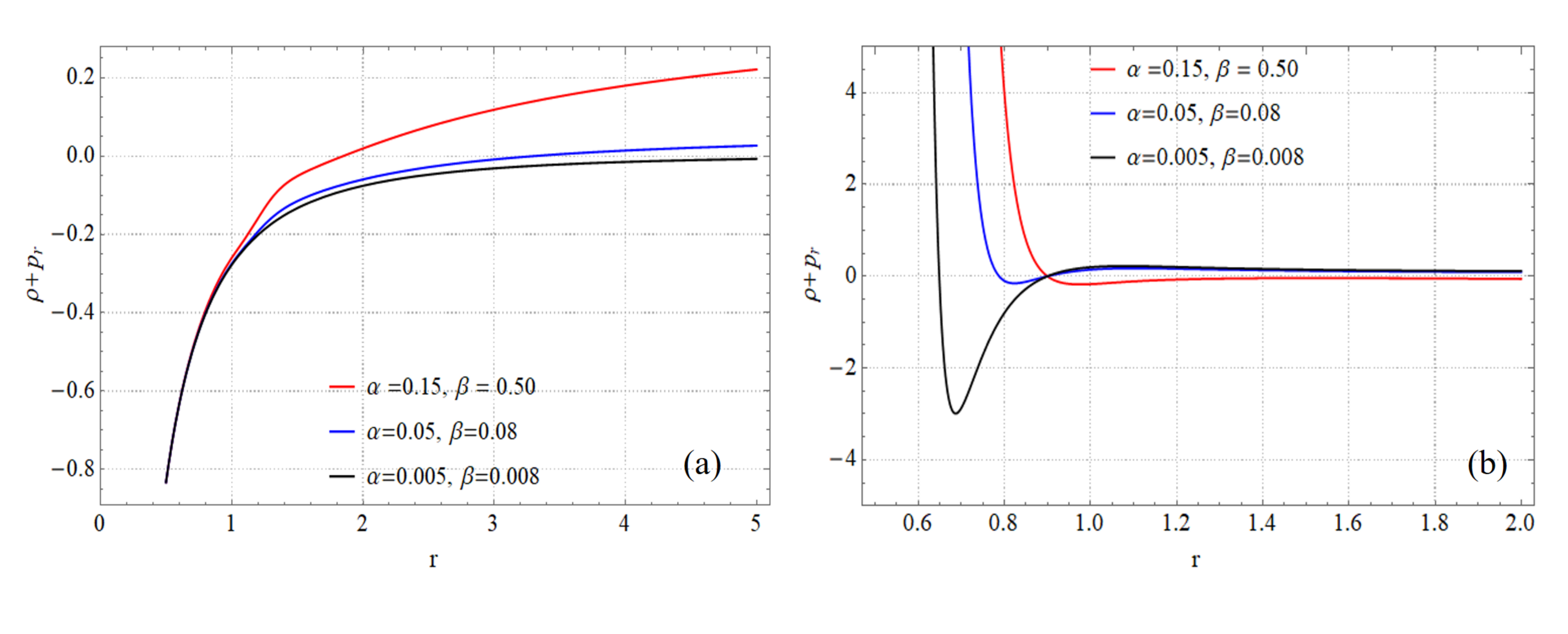}
   \caption{Profile of the first NEC term for different combinations of model parameters in arbitrary units with $R_c=1$ for (a) Case I (constant redshift function) and (b) Case II (variable redshift function).}\label{two}
\end{figure}
\begin{figure}[h]
   \includegraphics[width=\columnwidth]{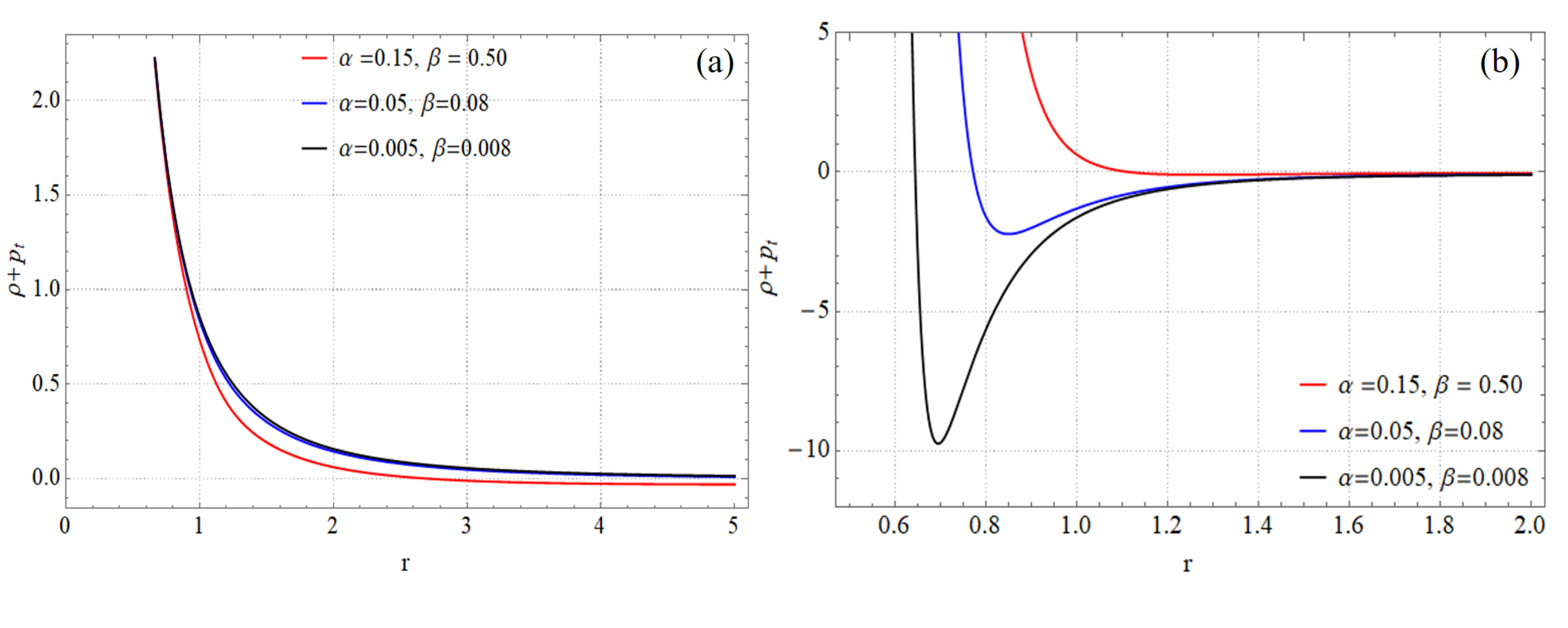}
   \caption{Profile of the second NEC term for different combinations of model parameters in arbitrary units with $R_c=1$ for (a) Case I (constant redshift function) and (b) Case II (variable redshift function).}\label{two}
\end{figure}
We also examined the SEC terms for both cases, and the results are shown in Fig. 4. The SEC is violated at the throat in Case I, and in Case II, the SEC holds with lower values of model parameters $\alpha$ and $\beta$. Moreover, the EoS parameters are $<-1$ in the throat region for both Cases, indicating a phantom-like profile of the source. Further, we have a positive anisotropy parameter in Case I for all combinations of model parameters, and in Case II, we have $\Delta>0$ for \textit{Model 1}. These results have been detailed in Tables 1 and 2.
\\
The above results demonstrate that we obtain a wormhole space-time satisfying the NEC in Model 1, supported by a source with a phantom-energy like source near the throat. Thus, wormholes satisfying the NEC at the throat can be formulated in the considered $f(R)$ model. Additionally, it can be seen that the model parameters $\alpha$ and $\beta$ have significant effects on the results. In Ref. \cite{Gogoi:2020ypn}, it has been highlighted that this parametric control (over $\alpha$ and $\beta$) is advantageous in realizing a model with solar-system and cosmological validity. The authors demonstrate that lower values of $\alpha$ and $\beta$ can make the model pass solar system tests \cite{Gogoi:2020ypn, Guo}. For wormholes in this $f(R)$ model, we highlight that higher values of model parameters (such as in \textit{Model 1}) can support wormhole space-times satisfying the NEC.
\begin{figure}[H]
   \includegraphics[width=\columnwidth]{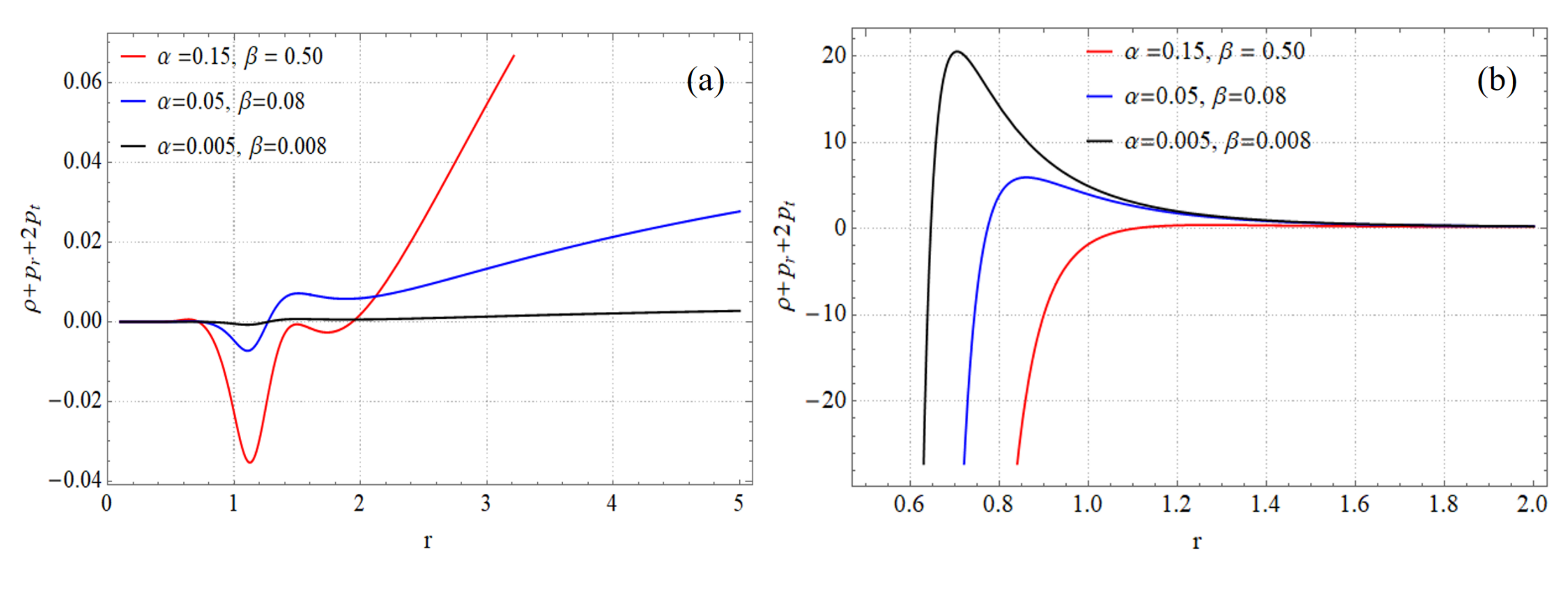}
   \caption{Profile of the SEC for different combinations of model parameters in arbitrary units with $R_c=1$ for (a) Case I (constant redshift function) and (b) Case II (variable redshift function).}\label{two}
\end{figure}
\begin{figure}[H]
   \includegraphics[width=\columnwidth]{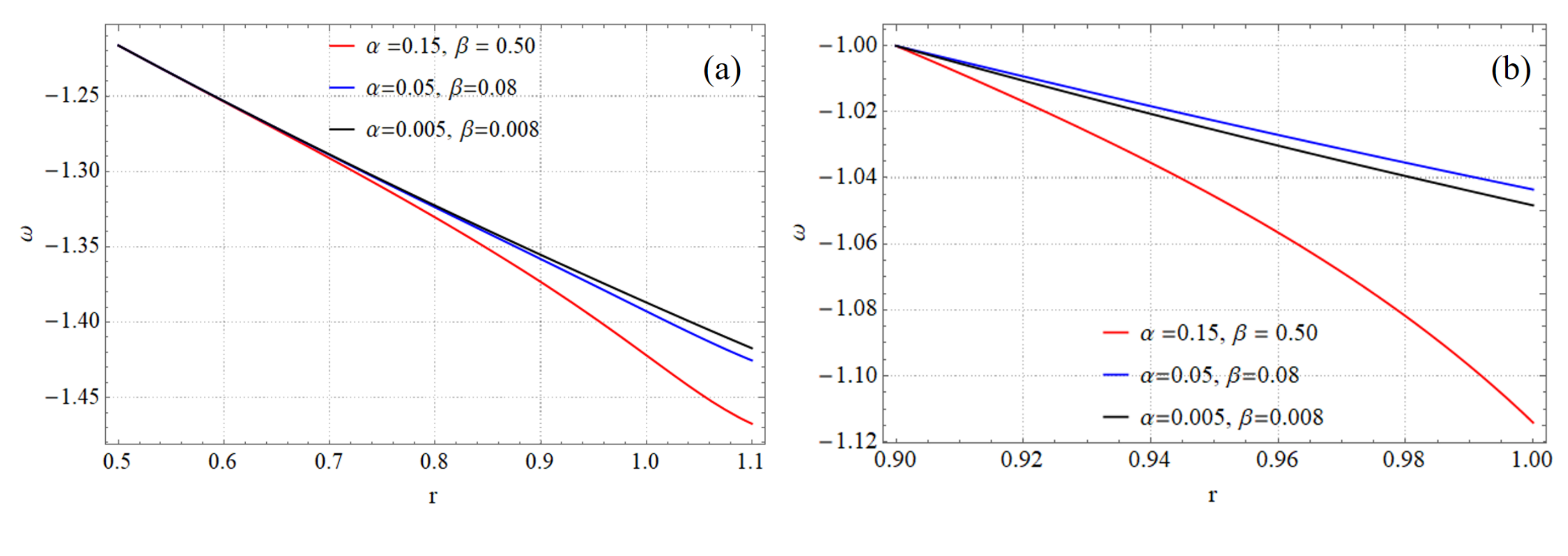}
   \caption{Profile of the EoS parameter for different combinations of model parameters in arbitrary units with $R_c=1$ for (a) Case I (constant redshift function) and (b) Case II (variable redshift function).}\label{two}
\end{figure}
\begin{figure}[H]
   \includegraphics[width=\columnwidth]{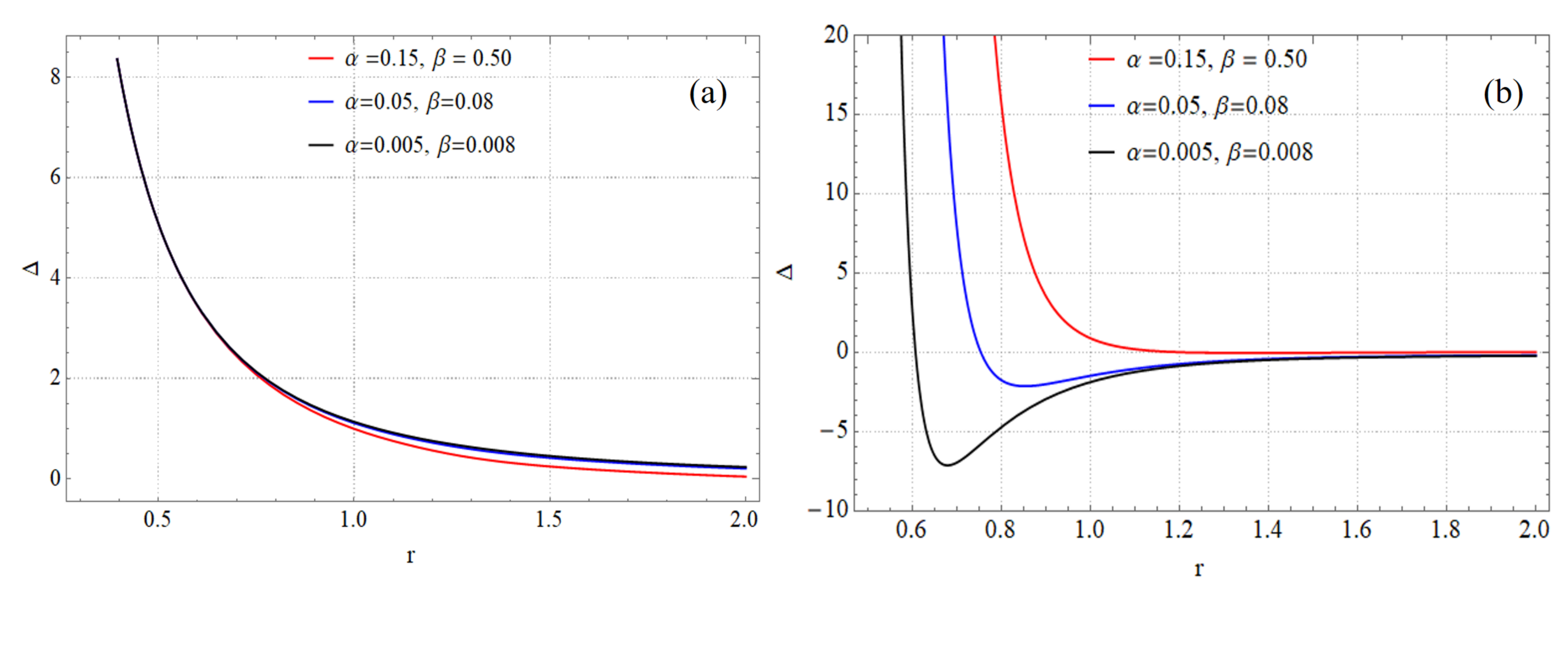}
   \caption{Profile of the anisotropy parameter for different combinations of model parameters in arbitrary units with $R_c=1$ for (a) Case I (constant redshift function) and (b) Case II (variable redshift function).}\label{two}
\end{figure}
Additionally, considering the generalized NEC \cite{capo2, capozziello2015} as described in Section \ref{sec:intro}:
\begin{equation}\label{eq:gnec}
    T_{\mu \nu} k^\mu k^\nu + k^\mu k^\nu \frac{\nabla_\mu \nabla_\nu F}{F} \geq 0,
\end{equation}
the second term on the LHS takes the following form with our choice of $f(R)$:
\begin{align}
    \frac{64 \alpha  R_c^5 r^{10} {b}^3 e^{\frac{2 b'}{R_c r^2}} \left(5 R_c^4 r^8-48 {b'}^4\right)-\pi  \beta  \left(16 {b'}^4+R_c^4 r^8\right)^3}{\left(16 R_c {b'}^4+R_c^5 r^8\right)^2 \left[e^{\frac{2 b'}{R_c r^2}} \left(16 \pi  {b'}^4+\pi  R_c^4 r^8 -4 \alpha  R_c^3 r^6 b' \right )-\pi  \beta  \left(16 {b'}^4+R_c^4 r^8\right)\right]} 
\end{align}
Here, one may set $\alpha$ and $\beta$ such that the inequality Eq. \ref{eq:gnec} holds, at least at the throat $r=r_0$. In our analyses, we select $\alpha$ and $\beta$ in ranges taking into account the viability of the considered $f(R)$ model. Our results highlight that although the generalized NEC Eq. \ref{eq:gnec} may be leveraged to develop solutions respecting the NEC, practical constraints on model parameters in our approach may lead to unstable solutions when standard matter sources are required to respect the NEC, as described next.
\\
With these insights, we analyzed the stability of the wormhole space-times using the generalized TOV equation \cite{PhysRev.55.374,tov2,tov3,tov4}:
\begin{equation}\label{eq:tov}
    -\frac{dp_{r}}{dr}-\frac{\epsilon^{'}(r)}{2}(\rho+p_{r})+\frac{2}{r}(p_{t}-p_{r})=0,
\end{equation}
where $\epsilon(r)=2\Phi(r)$. The three terms in the TOV equation represent the hydrostatic ($F_h$), gravitational ($F_g$), and anisotropic forces ($F_a$) in the space-time, respectively, and determine the equilibrium anisotropic mass distribution\cite{tov4}. 
\begin{equation}\label{28}
F_{\mathrm{h}}=-\frac{dp_{r}}{dr},\;\;\;\;\;\;\;\;F_{\mathrm{a}}=\frac{2}{r}(p_{t}-p_{r}), \;\;\;\;\;\;\;\;F_{\mathrm{g}}=-\frac{\epsilon^{'}}{2}(\rho+p_{r})
\end{equation}
The TOV equation must hold for spherically symmetric wormholes to be in equilibrium (stable configuration), and we checked the corresponding terms graphically in Fig. 5. It evident that $F_g$ vanishes for Case I ($\Phi'(r)=0$). The analyses show that stable configurations can be obtained only by decreasing $\alpha$ and $\beta$. Fig. 5 shows plots of $F_h$, $F_g$, and $F_a$ with $\alpha=0.005$ and $\beta=0.008$. Fig. 5(a) shows the equilibrium for the wormhole in Case I through the opposite behavior of $F_h$ and $F_a$, where the terms cancel each other to satisfy Eq. \eqref{eq:tov}. Similarly, for Case II, $F_h$ dominates over $F_g$ and $F_a$, and it is balanced by the combined effect of the $F_g$ and $F_a$. In the case of \textit{Model 1}, the TOV equation is not satisfied by the corresponding terms, rendering the configuration unstable. These results have been omitted here for conciseness, and presented in \ref{AppendixA}.
\begin{figure}[H]
   \includegraphics[width=\columnwidth]{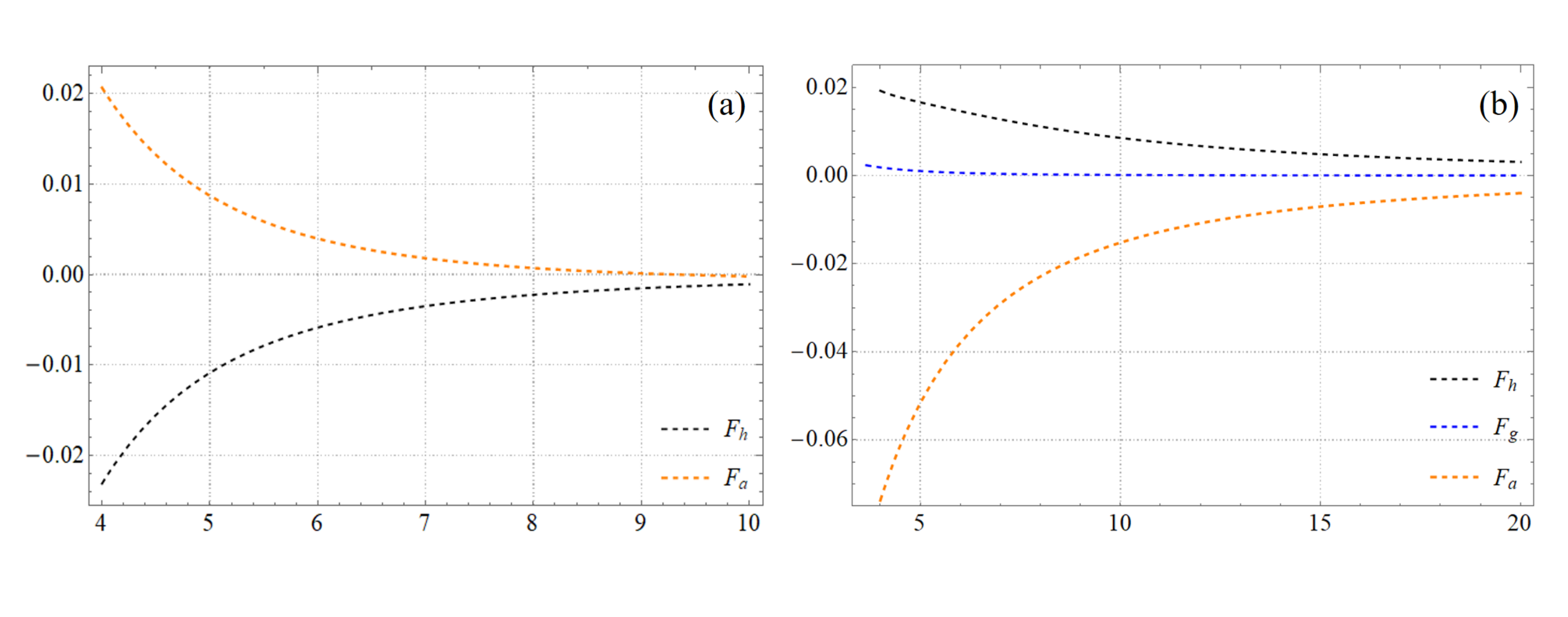}
    
    \caption{Equilibrium configurations for (a) Case I (constant redshift function) and (b) Case II (variable redshift function) with $\alpha=0.005$ and $\beta=0.008$.}\label{two}
\end{figure}

It is worth noting that we obtain a phantom-like profile of the source near the throat in both cases. Phantom wormholes satisfying the energy conditions in both GR and $f(R)$ gravity have been reported previously (for example in Ref. \cite{PhysRevD.71.084011} and Refs. \cite{phantommain, sahoo2018phantom}), and these solutions have important implications. While $\omega < -1$ is an exotic phenomenon considering the Standard Model, it cannot be ruled out considering current cosmological observations. Intriguing properties of phantom energy include a long-range repulsive force \cite{amendola1} manifesting as the accelerated expansion of the universe driven by an EoS with $\omega < -1$. An interesting candidate of phantom energy with a positive energy density as in our results is the \emph{axion}, conventionally described by an antisymmetric rank-3 tensor field, and it is of significant interest in string theory and supergravity. Moreover, we obtained a repulsive nature of the space-time geometries near the throat, which further support the anisotropic EoS in our results.\\
\vspace{0.8cm}
\begin{center}
\captionof{table}{Summary of results for Case I.}
{\begin{tabular}{|c|c|c|}
    \hline
    Term & Result & Interpretation \\
    \hline
    $\rho$ & $>0 \, \forall r$ in all Cases & WEC satisfied \\ 
    \hline
    $\rho+p_r$ & \makecell{\small $<0$ for  $r \in (0,1.8)$, \, $>0$ for $r \in (1.8,\infty)$\\with $\alpha = 0.15$, $\beta=0.50$ \\ \small $<0$ for  $r \in (0,3.3)$, \, $>0$ for $r \in (3.3,\infty)$ \\ with $\alpha = 0.05$, $\beta=0.08$ \\ \small$<0$ for  $r \in (0,7.3) $, \, $>0$ for $r \in (7.3,\infty)$ \\with $\alpha = 0.005$, $\beta=0.008$}& \makecell{Not satisfied} \\
    \hline
    $\rho+p_t$ & \makecell{\small $>0$ for $r \in (0,2.7)$, \, $<0$ for $r \in (2.7,\infty)$\\ with $\alpha = 0.15$, $\beta=0.50$ \\ \small $>0$ for $r \in (0,8.3)$, \, $<0$ for $r \in (8.3,\infty)$ \\ with $\alpha = 0.05$, $\beta=0.08$ \\ \small $\geq 0 \forall$ $r$ \\ with$\alpha = 0.005$, $\beta=0.008$ } & \makecell{Satisfied} \\
    \hline
    $\rho+p_r+2p_t$ & \makecell{$< 0$ for $r \in (0.7,1.9)$ and $\geq0$ otherwise \\ with $\alpha = 0.15$, $\beta=0.50$ \\ $< 0$ for $r \in (0.7,1.4)$ and $\geq0$ otherwise \\ with $\alpha = 0.05$, $\beta=0.08$\\ $< 0$ for $r \in (0.9,1.4)$ and $\geq0$ otherwise \\ with $\alpha = 0.005$, $\beta=0.008$} &\makecell{Not satisfied}  \\
    \hline
    $\omega$ & $<-1$ near throat for all Cases & Phantom-like source\\
    \hline
    $\Delta$ & $>0$ near the throat & \makecell{Repulsive geometry near throat} \\
    \hline
    \end{tabular}\label{tab:one}}
\end{center}
  
\begin{center}
\captionof{table}{Summary of results for Case II.}
{\begin{tabular}{|c|c|c|}
    \hline
    Term & Result & Interpretation \\
    \hline
    $\rho$ & \makecell{\small $>0$ for $r\in(0,1.1)$, $<0$ for $r\in(1.1,\infty)$ \\ with $\alpha=0.15$, $\beta=0.50$ \\ \small $>0$ for $r\in(0,0.7)$, $<0$ for $r\in(0.7,\infty)$ \\ with $\alpha=0.05$, $\beta=0.08$ \\ \small $>0$ for $r\in(0,0.6)$, $<0$ for $r\in(0.6,\infty)$ \\ with $\alpha=0.005$, $\beta=0.008$} & \makecell{WEC satisfied \\ in \textit{Model 1}} \\ 
    \hline
    $\rho+p_r$ & \makecell{\small $<0$ for $r \in (0.9,\infty)$, $>0$ for  $r \in (0,0.9)$\\with $\alpha = 0.15$, $\beta=0.50$ \\ \small $<0$ for  $r \in (0.8,0.9)$, $>0$ for $r \in (0,0.8) \cup (0.9,\infty)$ \\ with $\alpha = 0.05$, $\beta=0.08$ \\ \small$<0$ for  $r \in (0.6,0.9) $, $>0$ for $r \in (0,0.6) \cup (0.9,\infty)$ \\with $\alpha = 0.005$, $\beta=0.008$}& \makecell{NEC satisfied \\ in \textit{Model 1}} \\
    \hline
    $\rho+p_t$ & \makecell{\small $>0$ for $r \in (0,1.1)$, $<0$ for $r \in (1.1,\infty)$\\ with $\alpha = 0.15$, $\beta=0.50$ \\ \small $>0$ for $r \in (0,0.7)$, $<0$ for $r \in (0.7,\infty)$ \\ with $\alpha = 0.05$, $\beta=0.08$ \\ \small $>0$ for $r \in (0,0.6)$, $<0$ for $r \in (0.6,\infty)$ \\ with $\alpha = 0.005$, $\beta=0.008$ } & \makecell{NEC satisfied \\ in \textit{Model 1}} \\
    \hline
    $\rho+p_r+2p_t$ & \makecell{$< 0$ for $r \in (0,1.1)$ and $>0$ otherwise \\ with $\alpha = 0.15$, $\beta=0.50$ \\ $< 0$ for $r \in (0,0.7)$ and $>0$ otherwise \\ with $\alpha = 0.05$, $\beta=0.08$\\ $< 0$ for $r \in (0,0.6)$ and $>0$ otherwise \\ with $\alpha = 0.005$, $\beta=0.008$} &\makecell{SEC not satisfied \\ in \textit{Model 1}} \\
    \hline
    $\omega$ & $<-1$ near throat for all Cases & Phantom-like source\\
    \hline
    $\Delta$ & \makecell{$>0$ near the throat \\ with $\alpha = 0.15$, $\beta=0.50$}  & \makecell{Repulsive geometry \\ in \textit{Model 1}} \\
    \hline
    \end{tabular}\label{tab:two}}
\end{center}
The trends of the energy condition inequalities far outside the wormhole throat have not been discussed, since the throat connecting two asymptotically flat regions is of primary concern in our analyses. To this end, it is feasible to analytically cut off such solutions at some $r_c$ away from the throat, and connect it to an exterior space-time at a junction interface with a cut-off of the stress-energy at $r_c$. Considering surface stress at the junction, a thin-shell is obtained around the wormhole (see for example Ref.\cite{2016xxx}), and the junction acts as a boundary surface in the absence of surface stress. Physically, this approach corresponds to wormhole space-times in which some different stress-energy distribution becomes dominant at $r>r_c$. In summary, our results show that wormholes satisfying the energy conditions can be formulated in the $f(R)$ gravity model analyzed in this study, but the resulting space-times will be unstable. The key factors here are the model parameters $\alpha$ and $\beta$. Decreasing these parameters can make the model pass solar system tests (establishing the viability of the model), and we obtain wormholes with characteristic NEC violations in this scenario. Finally, we comment briefly on the amount of exotic matter in the stable configurations in \ref{AppendixA}.

\section{Conclusions}\label{sec:conc}
In this study, we analyzed properties of Morris-Thorne wormhole space-times in a recently proposed $f(R)$ gravity model. To the best of our knowledge, this is the first report on wormhole solutions in this $f(R)$ model. Our results show that traversable wormholes satisfying the energy conditions can be realized with suitable choices of metric functions and model parameters. However, the resulting configurations do not exhibit stable behavior, as demonstrated using the TOV equation. We analyzed possible wormhole space-times considering both constant and variable redshift, and our results show that well-constrained $\Phi(r)$ can aid in ameliorating energy condition violations. However, analyses reveal that the stable configurations (constrained via parameters in the $f(R)$ functional) violate the NEC at the throat. Although current observations do not yield evidence for wormholes in space-time, wormhole physics present important implications for fundamental issues in gravity and the Standard Model. Furthermore, in addition to numerical and analytical wormhole solutions, potentially observable signatures of wormholes such as quasinormal modes and shadows are of significant interest in light of recent advances in VLBI techniques. These issues have not been addressed adequately in literature in the context of metric $f(R)$ gravity, and are subjects of future research.

\appendix
\section{Expressions of the energy density and pressure components}\label{AppendixB}

Case I: The expressions for energy density ($\rho$), radial pressure ($p_r$), and transverse pressure ($p_t$) evaluated from Eqs. \eqref{generic1}-\eqref{generic3} are:

\begin{align}
\rho = \frac{r_o \left[ 1 - \beta e^{ \frac{-2 r_o}{R_c r^2 \left(1+r\right) \log \left(r_o + 1\right)} } \right] - \frac{ 4 {R_c}^3 r_o r^6 {\left( 1 + r_o \right)}^3 \alpha \log \left( 1 + r_o \right)^3}{16 \pi {r_o}^4 + {R_c}^4 \pi r^8 {\left( 1 + r \right)}^4 \log {\left(1+ r_o \right)}^4}}{r^2 \left(1+r\right) \log \left( 1 + r_o\right)}
\end{align}

\begingroup
\allowdisplaybreaks
\begin{align}
p_r&=- \frac{r_o \log \left(1+r\right)}{r^3 \log \left(1 + r_o \right)} \left[ 1 - \beta e^{ \frac{-2r_o}{ R_c r^2 \left(1 +r \right) \log \left(1 + r_o \right)}} \right. \\ \nonumber
&- \left. \frac{4 {R_c}^3 r_o r^6 \alpha \log \left( 1 + r_o \right)^3 {\left(1+r\right)}^3}{16 \pi {r_o}^4 + {R_c}^4 \pi r^8 {\left(1+r\right)}^4 \log {\left(1+ r_o \right)}^4} \right] - \left[ \frac{\beta e^{ \frac{-2r_o}{R_c r^2 \left(1+r\right) \log \left(1+r_o\right)}}}{R_c} \right. \\ \nonumber
&- \left. \frac{2 {R_c}^3 \alpha \left( {R_c}^4 - \frac{48 {r_o}^4}{r^8 {\left(1+r\right)}^4} \right) }{\pi {\left( {R_c}^4 + \frac{16 {r_o}^4}{r^8 {\left(1+r\right)}^4 \log {\left(1+r_o\right)}^4} \right)}^2}  \right] \left[ \frac{r_o \left(-r + \left(1+r\right) \log \left(1+r\right) \right)}{2 r^2 \left(1 +r\right) \log \left(1 + r_o \right)} \right] \\ \nonumber
&-\left[ \frac{64 {R_c}^3 {r_o}^3 r^{10} {\left(1+r\right)}^5 \alpha \log {\left(1+r_o\right)}^5}{\pi {\left(16 {r_o}^4 + {R_c}^4 r^8 {\left(1+r\right)}^4 \log {\left(1+r\right)}^4 \right)}^3} \right. \\ \nonumber
&\times \left. \left\{ 48 {r_o}^4 - 5 {R_c}^4 r^8 {\left(1+r\right)}^4 \log {\left(1+r_o\right)}^4\right\} \right. \\ \nonumber
&- \left.  \frac{\beta e^{ \frac{-2 r_o}{R_c r^2 \left(1+r\right) \log \left(1+r_o\right)}} }{{R_c}^2} \right] \left[1 - \frac{r_o \log \left(1+r\right)}{r \log \left(1+ r_o\right)} \right]
\end{align}
\endgroup

\begingroup
\allowdisplaybreaks
\begin{align}
p_t&= \frac{r_o\left[1-\beta e^{\frac{-2r_o}{R_c r^4\left(1+r\right)\log\left(1+r_c\right)}}-\frac{4 R_c^{3}r_or^6\left(1+r\right)^3\alpha\log\left(1+r_o\right)^3}{16\pi r_o^4+R_c^4\pi r^8\left(1+r\right)^4\log\left(1+r_o\right)^4}\right]}{2r^3\left(1+r\right)\log\left(1+r_o\right)} \\ \nonumber
&\times \Bigg[-r+\left(1+r\right)\log\left(1+r\right)\Bigg]- \left[1-\frac{r_o\log\left(1+r\right)}{r\log\left(1+r_o\right)}\right] \\ \nonumber
&\times \frac{1} {r^2}\left[ \frac{\beta e^{\frac{-2r_o}{R_c r^2\left(1+r\right)log\left(1+r_o\right)}}}{R_c }-\frac{2R_c^3\alpha\left(\frac{48r_o^4}{r_8\left(1+r\right)^4log\left(1+r_o\right)^4}\right)}{\pi\left(R_c^4+\frac{16r_o^4}{r^8\left(1+r\right)^4\log\left(1+r_o\right)^4}\right)^2} \right]
\end{align}
\endgroup

Case II: The expressions for energy density ($\rho$), radial pressure ($p_r$), and transverse pressure ($p_t$) evaluated from Eqs. \eqref{vrsfone}-\eqref{vrsfthree} are:

\begin{align}
\rho&=\frac{1}{2} \left[-\frac{\alpha}{\pi } R_c \cot ^{-1}\left(\frac{R_c^2 r^6}{4 r_0^2}\right)-\beta  R_c \left\{1-e^{\frac{2 r_0}{R_c r^3}}\right\}-\frac{2 r_0}{r^3}\right] +\frac{2 r_0}{r^3} \left[-\frac{\alpha}{\pi}  R_c \cot ^{-1}\left(\frac{R_c^2 r^6}{4 r_0^2}\right) \right. \nonumber \\
&- \left. \beta  R_c \left\{1-e^{\frac{2 r_0}{R_c r^3}}\right\}-\frac{2 r_0}{r^3}\right] + \frac{\sqrt{r_0}}{2 r^{\frac{3}{2}}} \left[1-\frac{r_0 \log \left(\frac{r}{r_0}\right)+r_0}{r}\right] \left[\frac{\alpha  R_c^3 r^9}{4 \pi  r_0^3 \left(\frac{R_c^4 r^{12}}{16 r_0^4}+1\right)}-\beta  e^{\frac{2 r_0}{R_c r^3}}+1\right] \nonumber \\
&- \left[1-\frac{r_0 \log \left(\frac{r}{r_0}\right)+r_0}{r}\right] \left[-\frac{\alpha  R_c^7 r^{24}}{32 \pi  r_0^8 \left(\frac{R_c^4 r^{12}}{16 r_0^4}+1\right)^2}+\frac{3 \alpha  R_c^3 r^{12}}{8 \pi  r_0^4 \left(\frac{R_c^4 r^{12}}{16 r_0^4}+1\right)} \right. \nonumber \\
&+ \left. \frac{2}{r} \left\{\frac{\alpha  R_c^3 r^9}{4 \pi  r_0^3 \left(\frac{R_c^4 r^{12}}{16 r_0^4}+1\right)}-\beta  e^{\frac{2 r_0}{R_c r^3}}+1\right\} - \frac{\sqrt{r_0}}{2 r^{\frac{3}{2}}} \left\{\frac{\alpha  R_c^3 r^9}{4 \pi r_0^3 \left(\frac{R_c^4 r^{12}}{16 r_0^4}+1\right)}-\beta  e^{\frac{2 r_0}{R_c r^3}}+1\right\} \right. \nonumber \\
&- \left. \frac{r_0 \log \left(\frac{r}{r_0}\right)}{2 r^2 \left(1-\frac{r_0 \log \left(\frac{r}{r_0}\right)+r_0}{r}\right)} \left\{\frac{\alpha  R_c^3 r^9}{4 \pi  r_0^3 \left(\frac{R_c^4 r^{12}}{16 r_0^4}+1\right)}-\beta  e^{\frac{2 r_0}{R_c r^3}}+1\right\} + \frac{\beta e^{\frac{2 r_0}{R_c r^3}}}{R_c}\right]
\end{align}

\begingroup
\allowdisplaybreaks
\begin{align}
p_r&=\frac{1}{2} \left[\frac{\alpha}{\pi} R_c \cot ^{-1}\left(\frac{R_c^2 r^6}{4 r_0^2}\right) + \beta R_c \left\{1-e^{\frac{2 r_0}{R_c r^3}}\right\} + \frac{2 r_0}{r^3}\right] - \frac{r_0}{r^3} \left[-\frac{\alpha}{\pi}  R_c \cot ^{-1}\left(\frac{R_c^2 r^6}{4 r_0^2}\right) \right. \nonumber \\
&- \left. \beta R_c \left\{1-e^{\frac{2 r_0}{R_c r^3}}\right\} - \frac{2 r_0}{r^3}\right] + \frac{1}{r^3} \left[r_0 \left\{-\log \left(\frac{r}{r_0}\right)\right\}-r_0\right] \left[-\frac{\alpha}{\pi}  R_c \cot ^{-1}\left(\frac{R_c^2 r^6}{4 r_0^2}\right) \right. \nonumber \\
&- \left. \beta R_c \left\{1-e^{\frac{2 r_0}{R_c r^3}}\right\} - \frac{2 r_0}{r^3}\right] +\frac{2 \sqrt{r_0}}{r^{\frac{3}{2}}} \left[1-\frac{r_0 \log \left(\frac{r}{r_0}\right)+r_0}{r}\right] \left[-\frac{\alpha}{\pi}  R_c \cot ^{-1}\left(\frac{R_c^2 r^6}{4 r_0^2}\right) \right. \nonumber \\
&- \left. \beta  R_c \left\{1-e^{\frac{2 r_0}{R_c r^3}}\right\} - \frac{2 r_0}{r^3}\right] - \left[1-\frac{r_0 \log \left(\frac{r}{r_0}\right)+r_0}{r}\right] \left[-\frac{\alpha  R_c^7 r^{24}}{32 \pi  r_0^8 \left(\frac{R_c^4 r^{12}}{16 r_0^4}+1\right)^2} \right. \nonumber \\
&+ \left. \frac{3 \alpha  R_c^3 r^{12}}{8 \pi  r_0^4 \left(\frac{R_c^4 r^{12}}{16 r_0^4}+1\right)} - \frac{r_0 \log \left(\frac{r}{r_0}\right)}{2 r^2 \left(1-\frac{r_0 \log \left(\frac{r}{r_0}\right)+r_0}{r}\right)} \left\{\frac{\alpha  R_c^3 r^9}{4 \pi  r_0^3 \left(\frac{R_c^4 r^{12}}{16 r_0^4}+1\right)}-\beta  e^{\frac{2 r_0}{R_c r^3}}+1\right\} \right. \nonumber \\
&+ \left. \frac{\beta  e^{\frac{2 r_0}{R_c r^3}}}{R_c}\right] + \left[1-\frac{r_0 \log \left(\frac{r}{r_0}\right)+r_0}{r}\right] \left[-\frac{\alpha  R_c^7 r^{24}}{32 \pi  r_0^8 \left(\frac{R_c^4 r^{12}}{16 r_0^4}+1\right)^2}+\frac{3 \alpha  R_c^3 r^{12}}{8 \pi  r_0^4 \left(\frac{R_c^4 r^{12}}{16 r_0^4}+1\right)} \right. \nonumber \\
&+ \left. \frac{2}{r} \left\{\frac{\alpha  R_c^3 r^9}{4 \pi  r_0^3 \left(\frac{R_c^4 r^{12}}{16 r_0^4}+1\right)}-\beta  e^{\frac{2 r_0}{R_c r^3}}+1\right\} - \frac{\sqrt{r_0}}{2 r^{\frac{3}{2}}} \left\{\frac{\alpha  R_c^3 r^9}{4 \pi  r_0^3 \left(\frac{R_c^4 r^{12}}{16 r_0^4}+1\right)}-\beta  e^{\frac{2 r_0}{R_c r^3}}+1\right\} \right. \nonumber \\
&- \left. \frac{r_0 \log \left(\frac{r}{r_0}\right)}{2 r^2 \left(1-\frac{r_0 \log \left(\frac{r}{r_0}\right)+r_0}{r}\right)} \left\{\frac{\alpha  R_c^3 r^9}{4 \pi  r_0^3 \left(\frac{R_c^4 r^{12}}{16 r_0^4}+1\right)}-\beta  e^{\frac{2 r_0}{R_c r^3}}+1\right\} + \frac{\beta  e^{\frac{2 r_0}{R_c r^3}}}{R_c}\right]
\end{align}
\endgroup

\begingroup
\allowdisplaybreaks
\begin{align}
p_t&=\frac{1}{2} \left[\frac{\alpha}{\pi} R_c \cot ^{-1}\left(\frac{R_c^2 r^6}{4 r_0^2}\right) + \beta R_c \left\{1-e^{\frac{2 r_0}{R_c r^3}}\right\} + \frac{2 r_0}{r^3}\right] - \frac{r_0}{r^3} \left[-\frac{\alpha}{\pi}  R_c \cot ^{-1}\left(\frac{R_c^2 r^6}{4 r_0^2}\right) \right. \nonumber \\
&- \left. \beta R_c \left\{1-e^{\frac{2 r_0}{R_c r^3}}\right\} - \frac{2 r_0}{r^3}\right] + \left[1-\frac{r_0 \log \left(\frac{r}{r_0}\right)+r_0}{r}\right] \left[-\frac{r_0^2}{4 r^4 \left(\frac{r_0}{r}\right)^{3/2}} \right. \nonumber \\
&- \left. \frac{ \sqrt{r} r_0^2 \log \left(\frac{r}{r_0}\right)}{4 r^3 \sqrt{r_0} \left(r_0 \left(-\log \left(\frac{r}{r_0}\right)\right)-r_0+r\right)}+\frac{\sqrt{r_0}}{2 r^{\frac{5}{2}}}\right] \left[-\frac{\alpha}{\pi} R_c \cot ^{-1}\left(\frac{R_c^2 r^6}{4 r_0^2}\right) \right. \nonumber \\
&- \left. \beta R_c \left\{1-e^{\frac{2 r_0}{R_c r^3}}\right\} - \frac{2 r_0}{r^3}\right] +\frac{r_0 \log \left(\frac{r}{r_0}\right)}{2 r^2} \left[-\frac{\alpha}{\pi} R_c \cot ^{-1}\left(\frac{R_c^2 r^6}{4 r_0^2}\right) - \beta R_c \left\{1-e^{\frac{2 r_0}{R_c r^3}}\right\} - \frac{2 r_0}{r^3}\right] \nonumber \\
&- \frac{1}{r}\left[1-\frac{r_0 \log \left(\frac{r}{r_0}\right)+r_0}{r}\right] \left[\frac{\alpha  R_c^3 r^9}{4 \pi r_0^3 \left(\frac{R_c^4 r^{12}}{16 r_0^4}+1\right)}-\beta  e^{\frac{2 r_0}{R_c r^3}}+1\right] + \left[1-\frac{r_0 \log \left(\frac{r}{r_0}\right)+r_0}{r}\right] \nonumber \\
&\times \left[-\frac{\alpha  R_c^7 r^{24}}{32 \pi r_0^8 \left(\frac{R_c^4 r^{12}}{16 r_0^4}+1\right)^2}+\frac{3 \alpha  R_c^3 r^{12}}{8 \pi  r_0^4 \left(\frac{R_c^4 r^{12}}{16 r_0^4}+1\right)}+\frac{2}{r} \left\{\frac{\alpha  R_c^3 r^9}{4 \pi  r_0^3 \left(\frac{R_c^4 r^{12}}{16 r_0^4}+1\right)}-\beta  e^{\frac{2 r_0}{R_c r^3}}+1\right\} \right. \nonumber \\
&- \left. \frac{\sqrt{r_0}}{2 r^{\frac{3}{2}}} \left\{\frac{\alpha  R_c^3 r^9}{4 \pi  r_0^3 \left(\frac{R_c^4 r^{12}}{16 r_0^4}+1\right)}-\beta  e^{\frac{2 r_0}{R_c r^3}}+1\right\} - \frac{r_0 \log \left(\frac{r}{r_0}\right)}{2 r^2 \left(1-\frac{r_0 \log \left(\frac{r}{r_0}\right)+r_0}{r}\right)} \left\{\frac{\alpha  R_c^3 r^9}{4 \pi  r_0^3 \left(\frac{R_c^4 r^{12}}{16 r_0^4}+1\right)} \right. \right. \nonumber \\
&- \left. \left. \beta  e^{\frac{2 r_0}{R_c r^3}}+1\right\} + \frac{\beta  e^{\frac{2 r_0}{R_c r^3}}}{R_c}\right]
\end{align}
\endgroup

\section{Stability analysis and \emph{amount} of exotic matter}\label{AppendixA}

\renewcommand{\thefigure}{A\arabic{figure}}
\setcounter{figure}{0}

\begin{figure}[htp]
   \subfloat[$F_h$]{\label{rev}
      \includegraphics[width=0.5\textwidth]{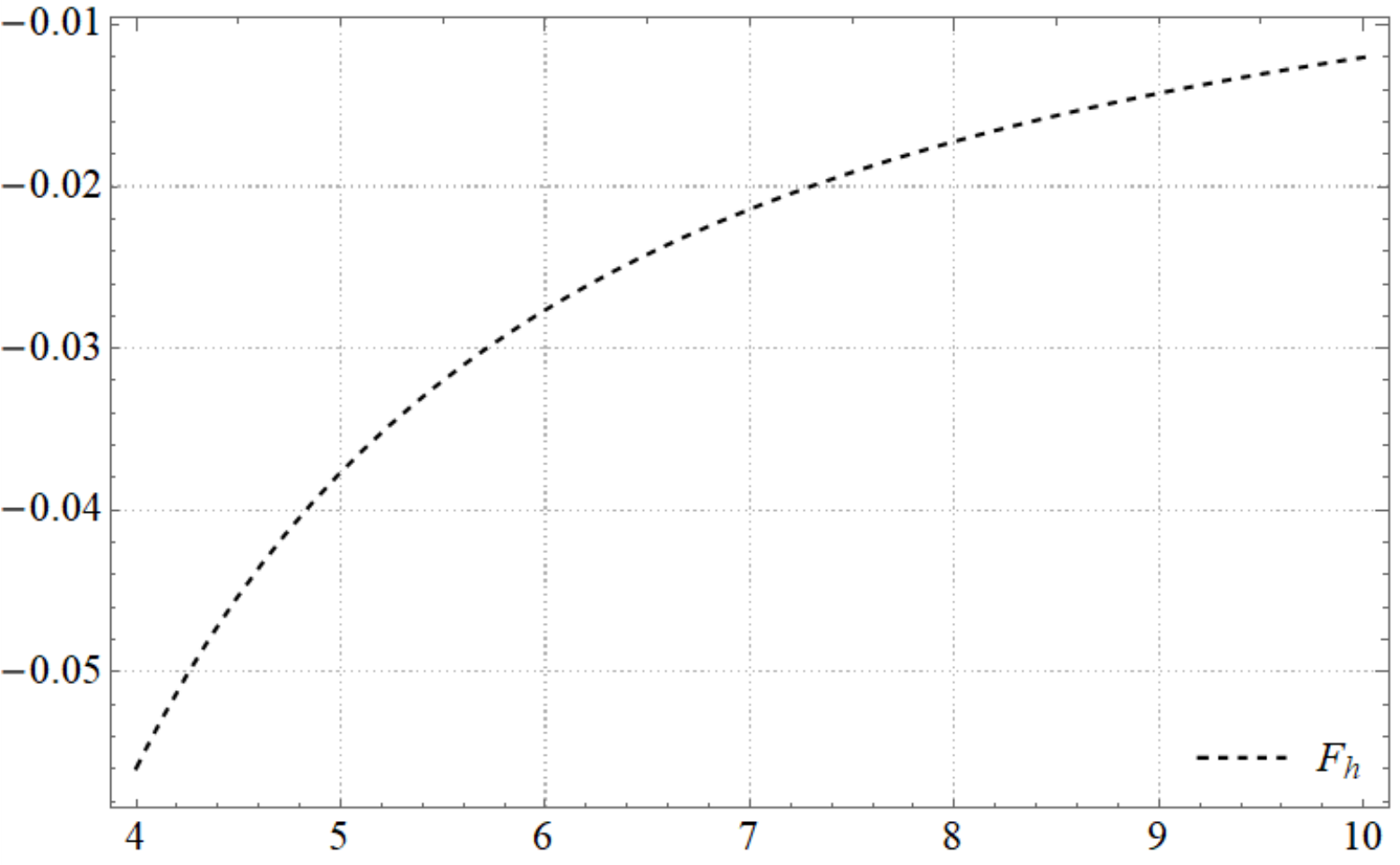}}
~
   \subfloat[$F_a$]{\label{rev_sol}
      \includegraphics[width=0.5\textwidth]{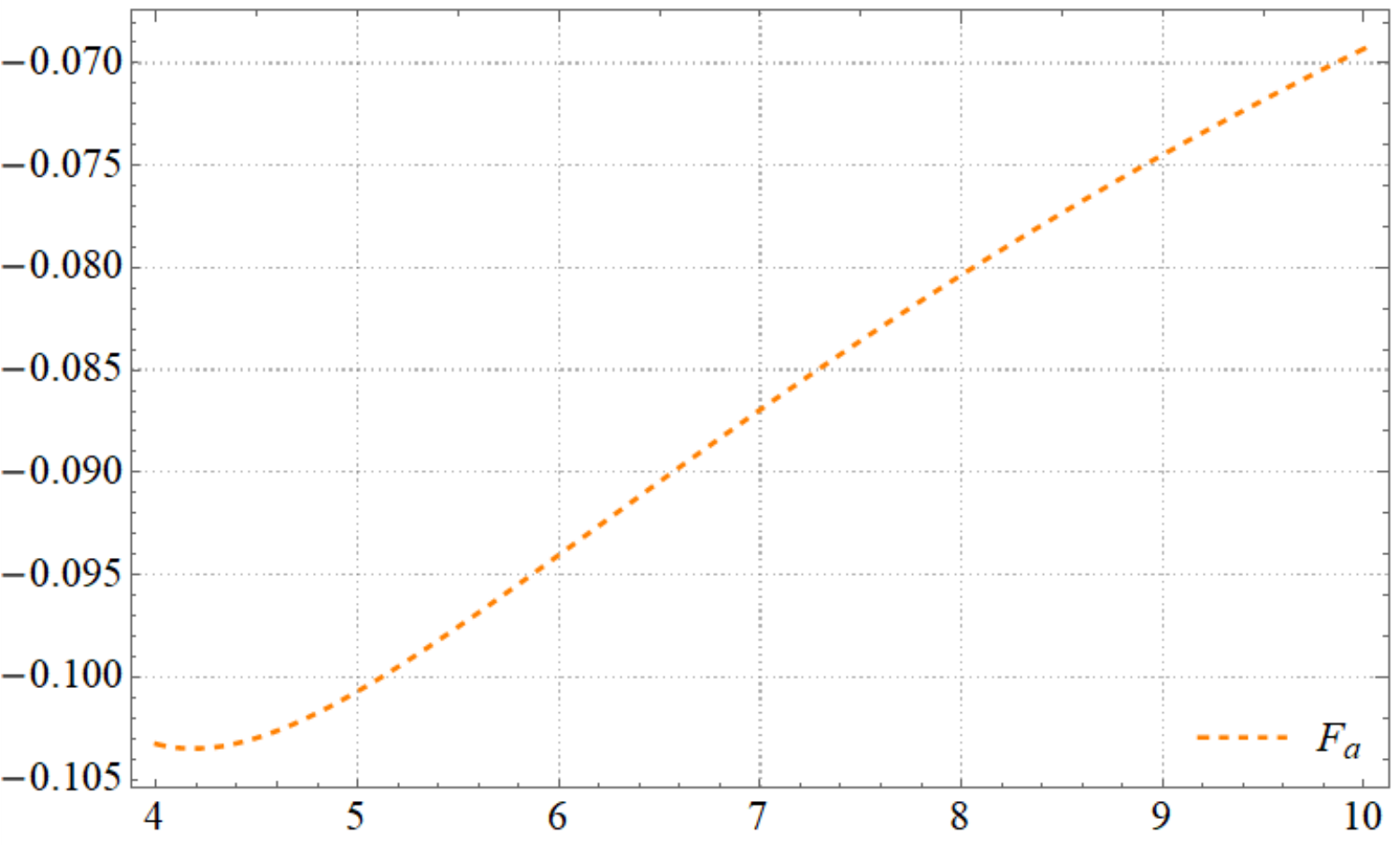}}

   \caption{Profile of the (a) hydrostatic and (b) anisotropic forces for Case I with $\alpha=0.15$, $\beta=0.05$.}\label{bs1}
\end{figure}

\begin{figure}[htp]
   \subfloat[$F_h$]{\label{rev}
      \includegraphics[width=0.32\textwidth]{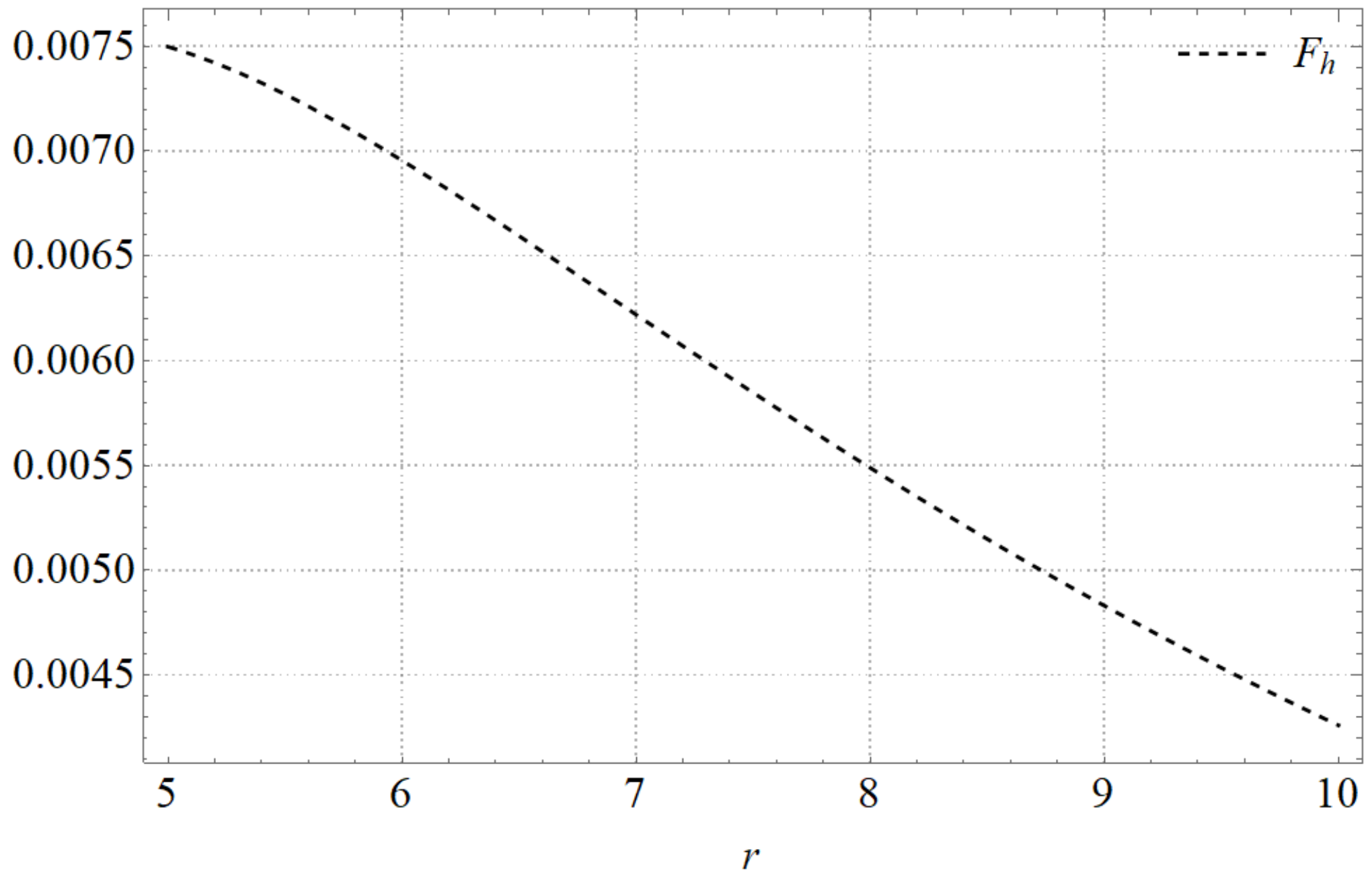}}
~
   \subfloat[$F_g$]{\label{rev_sol}
      \includegraphics[width=0.32\textwidth]{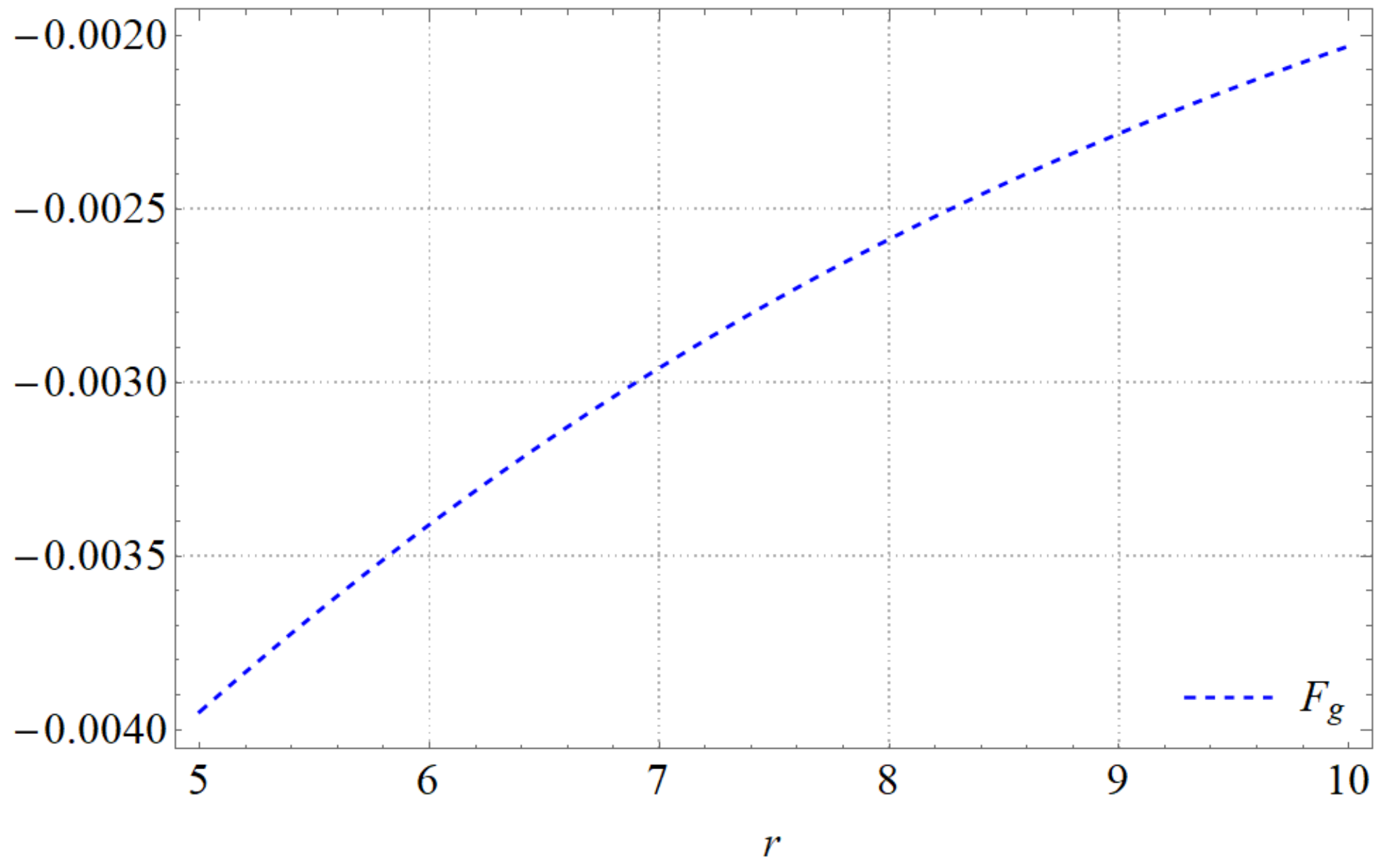}}
~
   \subfloat[$F_a$]{\label{rev_sol}
      \includegraphics[width=0.32\textwidth]{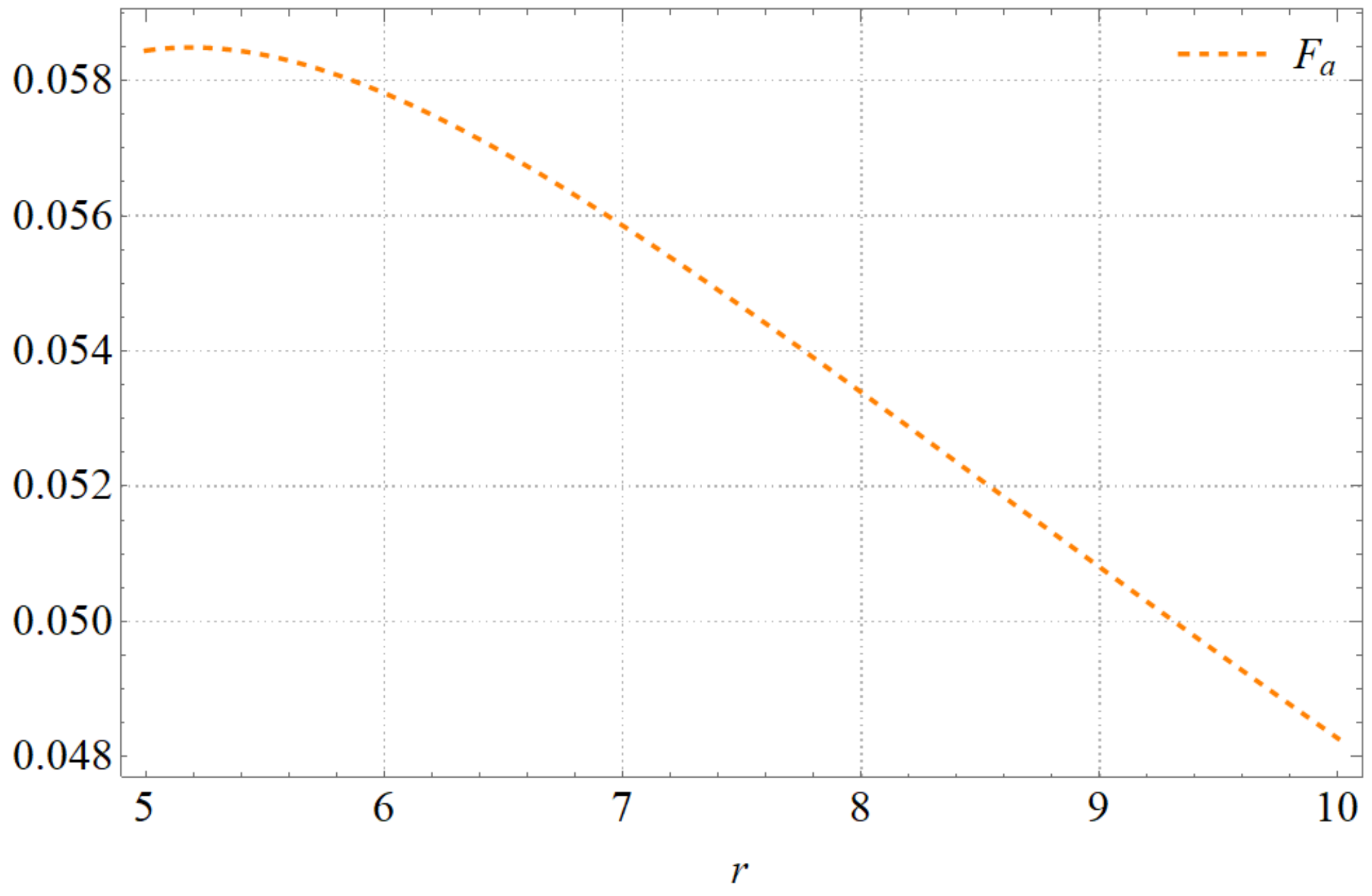}}

   \caption{Profiles of the (a) hydrostatic, (b) gravitational, and (c) anisotropic forces for Case II with $\alpha=0.15$, $\beta=0.05$.}\label{bs2}
\end{figure}

\subsection*{Amount of exotic matter near the throat}

\begin{figure}[htp]
   \subfloat[Case I]{\label{rev}
      \includegraphics[width=0.45\textwidth]{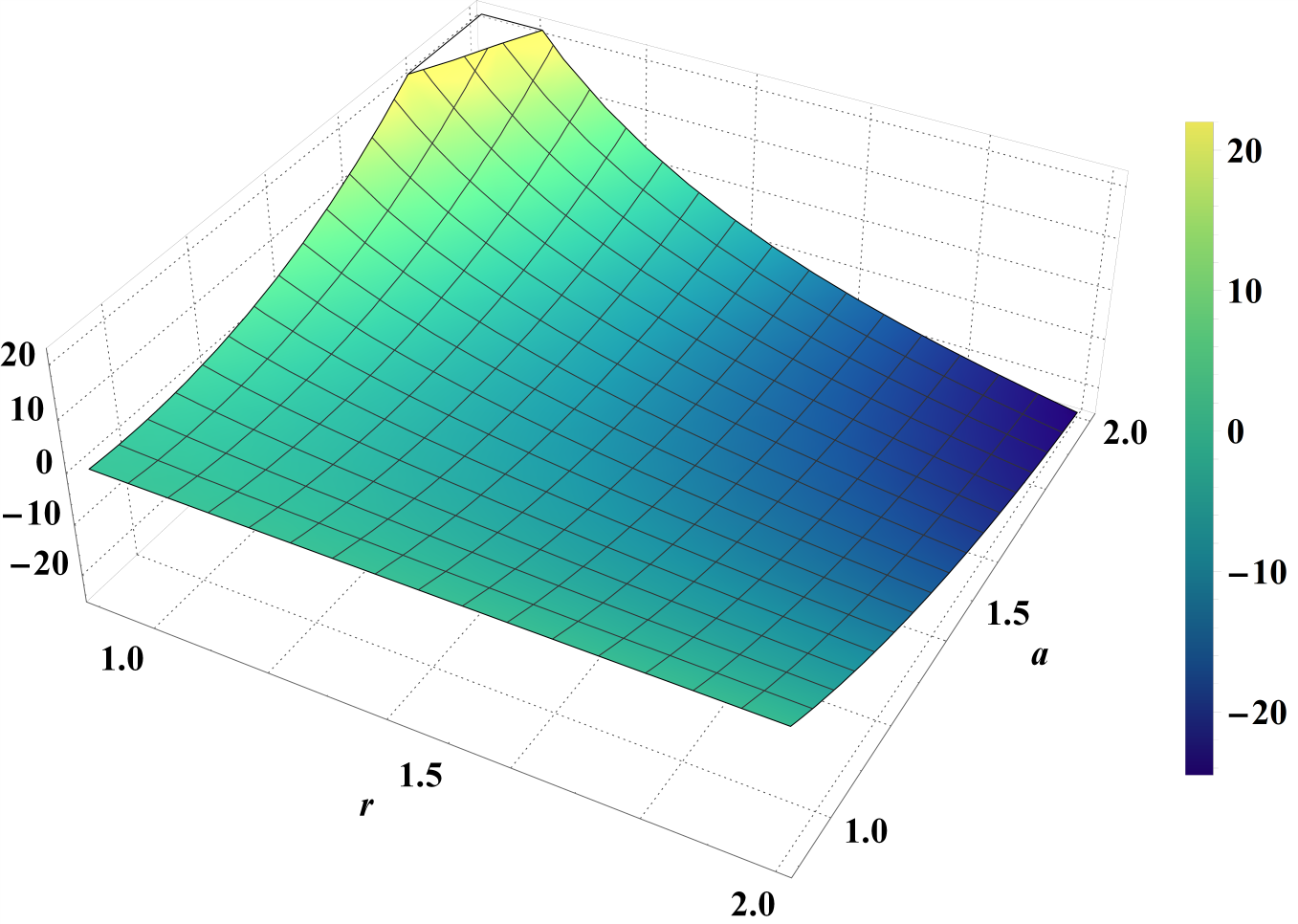}}
~
   \subfloat[Case II]{\label{rev_sol}
      \includegraphics[width=0.45\textwidth]{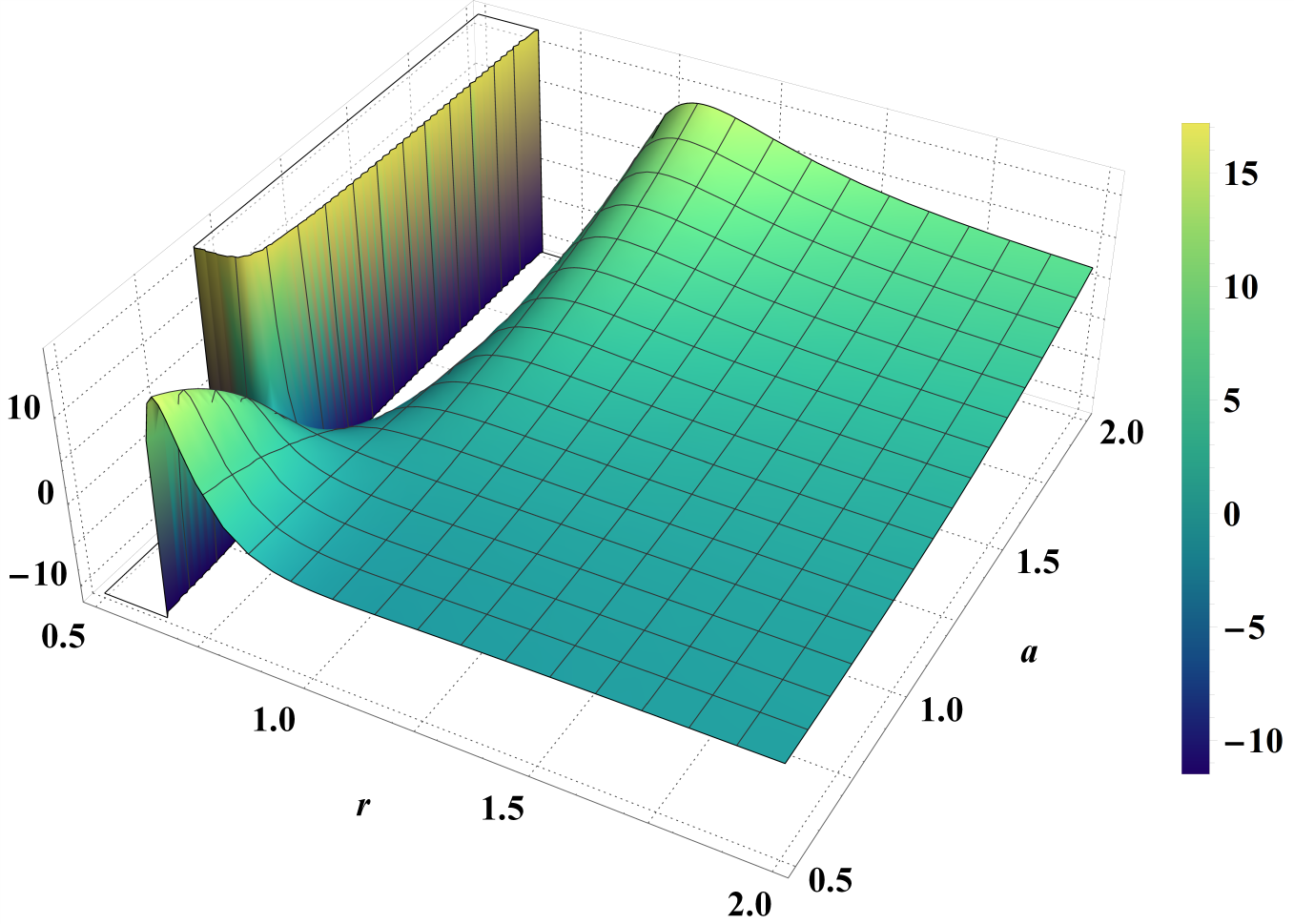}}
   \caption{Evolution of the volume integral quantifier with $r$ and $a$ for the stable configurations.}\label{bs3}
\end{figure}
The averaged null energy condition $\int_{\lambda_1}^{\lambda_2} T_{ij}k^ik^j d \lambda \geq 0$ can be evaluated to check for energy condition violations along the radial direction. However, since this is a line integral, it does not provide useful information regarding the amount of energy condition violating matter. A more generalized volume integral, called the volume integral quantifier\cite{viq1,viq2,viq3}, can be used to estimate the amount of exotic matter in spherically symmetric space-times.
\begin{equation}
    I_v = \oint [\rho+p_r] dV = 8\pi\int_{r_0}^{a}(\rho+p_r)r^2 dr
\end{equation}
Here, as described in Sec. 4, it is assumed that the wormhole space-time is matched with an exterior metric with a cut-off of the stress energy tensor at some $r=a$. As $a \rightarrow r_0$, $I_v \rightarrow 0$ implies that wormholes can be constructed with arbitrarily small quantities of NEC violating matter\cite{viq1,viq2}. Fig. A3 shows plots of $I_v$ for the stable configurations in Cases I and II, and it can be verified that as $a \rightarrow r_0$, $I_v \rightarrow 0$.

\printbibliography
\end{document}